\documentclass{article}

\usepackage{arxiv}

\usepackage{cite}
\usepackage{amsmath,amssymb,amsfonts}
\usepackage{algorithmic}
\usepackage{graphicx}
\usepackage{textcomp}
\usepackage{xcolor}
\usepackage{orcidlink}
\usepackage{import} 
\usepackage[binary-units]{siunitx}
\usepackage[ruled,vlined,linesnumbered]{algorithm2e}
\usepackage{subcaption}
\usepackage[square, numbers]{natbib}
\usepackage{csquotes}

\def\BibTeX{{\rm B\kern-.05em{\sc i\kern-.025em b}\kern-.08em
		T\kern-.1667em\lower.7ex\hbox{E}\kern-.125emX}}
	
\DeclareMathOperator*{\argmax}{arg\,max}
\DeclareMathOperator*{\argmin}{arg\,min}

\title{Multicast-partitioning in Time-triggered Stream Planning for Time-Sensitive Networks}

\author{\orcidlink{0000-0002-5146-0184}Heiko~Geppert\\
	IPVS\\
	University of Stuttgart\\
	Germany\\
	\texttt{heiko.geppert@uni-stuttgart.de} \\
	\And
	\orcidlink{0000-0002-3470-7712}Frank~D\"urr\\
	IPVS\\
	University of Stuttgart\\
	Germany\\
	\texttt{frank.duerr@uni-stuttgart.de} \\
	\And
	Simon Naß\\
	IPVS\\
	University of Stuttgart\\
	Germany\\
	\And
	\orcidlink{0000-0001-8986-8241}Kurt~Rothermel\\
	IPVS\\
	University of Stuttgart\\
	Germany\\
	\texttt{kurt.rothermel@uni-stuttgart.de} \\
}

\renewcommand{\shorttitle}{Multicast-partitioning for TSN}

\hypersetup{ 
	pdftitle={Multicast-partitioning in Time-triggered Stream Planning for Time-Sensitive Networks},
	pdfsubject={cs.NI},
	pdfauthor={Heiko Geppert et al.},
	pdfkeywords={multicast, time-sensitive networking, scheduling},
}

\begin{document}
	\maketitle
	
	
	\begin{abstract}
		
Multicast allows sending a message to multiple recipients without having to create and send a separate message for each recipient.
This preserves network bandwidth, which is particularly important in time-sensitive networks.
These networks are commonly used to provide latency-bounded communication for real-time systems in domains like automotive, avionics, industrial internet of things, automated shop floors, and smart energy grids.
The preserved bandwidth can be used to admit additional real-time messages with specific quality of service requirements or to reduce the end-to-end latencies for messages of any type.
However, using multicast communication can complicate traffic planning, as it requires free queues or available downstream egress ports on all branches of the multicast tree.
In this work, we present a novel multicast partitioning technique to split multicast trees into smaller multicast or unicast trees.
This allows for a more fine-grained trade-off between bandwidth utilization and traffic scheduling difficulty. 
Thus, schedulability in dynamic systems can be improved, in terms the number of admitted streams and the accumulated network throughput.
We evaluated the multicast partitioning on different network topologies and with three different scheduling algorithms.
With the partitioning, 5-15\% fewer streams were rejected, while achieving 5-125\% more network throughput, depending on the scheduling algorithm.
		
	\end{abstract}


\section{Introduction}

Real-time applications such as automotive systems, avionics, industrial automation, automated shop floors, smart energy grids, and healthcare \cite{Schweissguth2017,Nasrallah2019}, are designed to process data and respond within bounded time, thus ensuring predictable and deterministic behavior.
To meet these requirements in networked systems, real-time networks facilitate the timely and reliable exchange of data between interconnected devices, ensuring bounded latency, minimal jitter, and high synchronization accuracy. 

Ethernet is the dominant network technology for wired consumer networks, but its fast transmission speed and the ability to mix real-time traffic with best-effort traffic to reduce hardware and wiring made Ethernet attractive for real-time networking as well.
The Institute of Electrical and Electronics Engineers (IEEE) has reacted to this demand with a set of standards known as Time-Sensitive Networking (TSN), which specify real-time capabilities on top of the classical Ethernet.
Especially the \enquote{Enhancements for Scheduled Traffic} (IEEE 802.1Qbv \cite{IEEE802.1Qbv-2015}), which add the Time Aware Shaper (TAS), enable the precise planning of time-triggered traffic and allow for bounded end-to-end network delay.
This is done by a gating mechanism at the egress port queues.
A gate control list defines when and how long each gate is opened, so that frames in the corresponding egress queue can be forwarded or halted.
The scheduling synthesis itself is inherently complex as strict timing guarantees must be upheld with only limited network resources.
Therefore, the investigation of algorithms and techniques to create schedules for TSN networks has received much attention in research \cite{Stueber2023}.

Multicast is an essential Ethernet -- and also TSN -- feature that allows a single sender to communicate with multiple receivers.
It reduces network resource utilization by sending frames along a multicast tree and duplicating them only at branching points.
This makes efficient use of bandwidth by sending frames only once over each physical link, rather than sending multiple copies for each downstream receiver.

\begin{figure}
	\centering
	\includegraphics[width=0.7\linewidth]{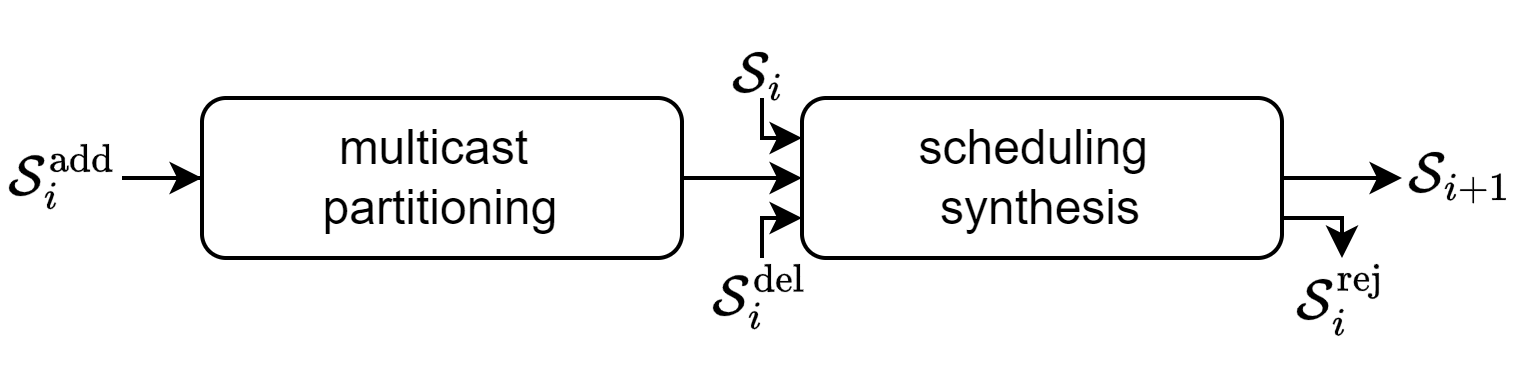}
	\caption{Two-phase process of multicast partitioning and scheduling. $\mathcal{S}$ denotes a set of streams: to be added ($\mathcal{S}_i^\text{add}$) to the system, removed ($\mathcal{S}_i^\text{del}$) from the system, that were rejected ($\mathcal{S}_i^\text{rej}$), or the admitted streams before ($\mathcal{S}_i$) and after ($\mathcal{S}_{i+1}$) the scheduling update.}
	\label{fig:two-step-process}
\end{figure}

The scheduling synthesis in TSN with multicast has received some attention in the past \cite{Steiner2010,Santos2019}.
Thereby, different optimizations have been investigated, including joined-routing-and-scheduling \cite{Schweissguth2020}, topology pruning, and stream clustering \cite{Li2021}.
Our proposed optimization works orthogonal to these techniques.
We consider the scheduling synthesis as a two phase workflow (cf. Figure~\ref{fig:two-step-process}).
First, the new streams' multicast trees are partitioned.
Then, in the second phase, the modified streams are forwarded to the scheduling algorithm.

The idea of our partitioning is that a multicast stream can be implemented as multiple smaller multicast or unicast streams.
The granularity of the partitioning lies on a spectrum, where the two extremes are well known.
On one extreme, a single multicast tree is used to reach all destinations of a stream.
This minimizes the bandwidth consumption of the stream.
However, scheduling many large multicast trees can be difficult.
It either requires the availability of all downstream egress ports simultaneously so that a frame arriving at a bridge can be forwarded immediately (required by no-wait schedules \cite{Duerr2016}), or egress queues must be available to buffer the frame.
In both cases, the scheduling becomes increasingly difficult with larger multicast trees.
The other extreme is to replace each multicast stream with a unicast stream to each destination \cite{Nie2022}.
In this case, the bandwidth requirements of such a partitioning would place a tremendous strain on the network, but the scheduling is much more flexible.

In this paper, we therefore present a novel multicast partitioning approach that splits multicast trees at an intermediate point between these two extremes. 
By splitting multicast trees with widely spread destinations into smaller multicast or even unicast trees, we improve the scheduling success while maintaining as many bandwidth utilization benefits as feasible.
The partitioning is based on a node distance threshold in hops which decides when and where to split a multicast tree, i.e., it regulates the partitioning granularity.
This paper explores the spectrum between the two extremes.
In addition, we study the interaction between partitioning and scheduling algorithms.

The remainder of the paper is structured as follows:
Related work is discussed in Section~\ref{sec:related-work}.
We specify our system model in Section~\ref{sec:system-model}, followed by a formal description of the scheduling synthesis and optimization goal in Section~\ref{sec:problem-statement}.
In Section~\ref{sec:multicast-partitioning}, we present our multicast partitioning approach and evaluate it in Section~\ref{sec:eval}.
Finally, we conclude our paper in Section~\ref{sec:conclusion}.


\section{Related Work}
\label{sec:related-work}

In the following, we provide a brief overview on TSN scheduling, especially in the context of multicast communication.
We particularly highlight the algorithms that we used in our evaluation for the workflow's second phase, or that were influential for our implementation.

Scheduling time-triggered traffic in TSN has been researched extensively in the past decades \cite{Stueber2023}.
The offline scheduling synthesis in smaller networks via mathematical solvers, in particular, Integer Linear Programming (ILP), Constraint Programming (CP), and Satisfiability Modulo Theory has been investigated in depth \cite{Santos2019,Steiner2010,Vlk2021}.
These solver-based approaches guarantee optimal results based on the problem definition, like minimal make span, at the cost of potentially high runtimes.

Several scheduling approaches specifically focus on multicast communication, while for others, it is mentioned that the scheduler can be extended to cover multicast streams \cite{Craciunas2016a,Pop2016,Bujosa2022}, but only unicast streams are presented in detail.
Schweissguth et al. investigate joined routing and scheduling (JRAS) for multicast streams using ILP \cite{Schweissguth2020}.
Li et al. propose topology pruning and stream clustering to reduce ILP runtimes \cite{Li2021}.
They show that JRAS reduced the makespan, while the clustering reduced the runtime.
The topology pruning reduces the number of variables and constraints in the ILP without reducing the solution space, reducing the runtime further.
Its impact is limited to topologies with \enquote{prunable} network parts and only comes into effect when using JRAS.
The CP implementation used in our evaluation was influenced by both ILP definitions.
However, as our focus lies on multicast partitioning, i.e., the first phase in our workflow, we did not implement JRAS or topology pruning for our second phase.

In the following we discuss two scalable scheduling approaches that we employed to evaluate our multicast partitioning.
The greedy approach from Raagaard et al., uses a computationally cheap heuristic for online TSN scheduling \cite{Raagaard2017}. 
The streams are scheduled iteratively, whereby frame transmissions are planned in an as-soon-as-possible fashion.
Planned transmissions are not changed even if they prevent later streams from being scheduled.
This makes the scheduling very fast, but at the cost of possibly higher stream rejections.
By only considering the shortest path for each stream the solution space stays manageable, while queuing allows for sufficient degrees of freedom.
The original approach does not include multicast streams, but can be extended trivially.

The second approach by Falk et al. uses conflict-graph-based scheduling \cite{Falk2020}.
The conflict graph models the (partial) solution space of the scheduling problem, which implicitly offers mechanisms to limit memory needs and complexity.
In addition, the graph structure enables the use of efficient graph algorithms such as the Greedy Flow Heap heuristic (GFH) to create schedules \cite{Falk2022}.
In our evaluation we used the improved version of the conflict-graph-based approach \cite{Geppert2024}, but extended it to work with multicast streams.


\section{System Model}
\label{sec:system-model}

This section discusses our system architecture.
We assume a real-time communication network $\mathcal{N}$ modeled as an undirected graph $\mathcal{N}(\mathcal{V},\mathcal{E})$.
The vertices $\mathcal{V}$ denote real-time bridges and end devices, while $\mathcal{E}$ models the links between them.
Each end device is connected to one bridge.

The bridges implement a store-and-forward behavior, i.e., a frame must be received completely bore it can be forwarded.
In the TAS, each bridge implements eight queues per egress port and frames are assigned to a queue based on the priority code point in the Ethernet header \cite{IEEE802.1Qbv-2015}.
A gate control list specifies when frames are forwarded from a bridge based on a precise schedule of gate openings and closings.
The bridges are synchronized using a protocol like, for example, gPTP \cite{IEEE802.1AS-2020} to a time granularity of \SI{1}{\micro\second} to enable coordinated scheduling.
The whole network is managed by a central network controller~\cite{IEEE802.1Qcc-2018}.

The communication in $\mathcal{N}$ is performed via cyclic, isochronous time-triggered streams $\mathcal{S}$ \cite{IECSC65C/WG18}.
A stream $s \in \mathcal{S}$ represents an unicast or multicast communication and consists of a source, destination(s), period, frame size, and deadline.
The source denotes the end device from which the stream originates.
A stream can have one or multiple destinations, which are all end devices.
The period describes the stream's cycle time, i.e., the time between the release of two consecutive frames of a stream.
The frame size is the number of bytes transmitted in each cycle and the deadline denotes the maximum end-to-end latency.
Without loss of generality we assume that period = deadline.
The scheduling synthesis is performed on a per-stream level.

The set of streams $\mathcal{S}$ changes over time: new streams can attempt to join the system, admitted streams can leave the system or be reconfigured to improve resource utilization.
These dynamics are modeled as a set of continuous iterations $\mathcal{I}$.
The streams that are admitted in iteration $i+1 \in \mathcal{I}$ are denoted as:

\begin{equation}
	\mathcal{S}_{i+1} = \bigl (\mathcal{S}_i \setminus \mathcal{S}^\text{del}_i \bigr ) \cup \bigl (\mathcal{S}^\text{add}_i \setminus \mathcal{S}^\text{rej}_i \bigr )
\end{equation}

Thereby, $\mathcal{S}_i$ is the set of streams that were admitted in the previous iteration $i$.
The set $\mathcal{S}^\text{add}_i$ collects the streams that attempt to join during iteration $i$ with the earliest admission in $i+1$.
$\mathcal{S}^\text{del}_i$ contains the streams that are admitted in $i$, but will leave the system at the transition to iteration $i+1$.
$\mathcal{S}^\text{rej}_i$ consists of the streams from $\mathcal{S}^\text{add}_i$ that cannot be admitted in iteration $i+1$ and are therefore rejected.


\section{Problem Statement}
\label{sec:problem-statement}

In the following, we first describe the requirements for a valid TSN schedule, followed by our optimization goal.

\subsection{TSN Scheduling Synthesis}

A TSN schedule describes a timetable for every frame of all time-triggered streams in $\mathcal{S}$.
For the schedule to be valid, the following properties must be fulfilled: valid release time, deadlines, precedence constraints, frame isolation, and mutual exclusive transmissions.

Each frame has a known \emph{release time} relative to the start of the hyper cycle.
As the sender cannot send a frame earlier than its release time, the schedule must offer a transmission slot after the specified release time.
The release time can also be defined by an expected release time with a maximum release time jitter.
Then the worst case release time is to be considered.
The \emph{deadlines} of all frames must be met in order to protect the real-time properties of the application.
Thus, the schedule must ensure that each frame of an admitted stream arrives at all destinations no later than specified in its deadline.
To ensure correct link-level \emph{precedence}, a bridge must receive the whole frame and process it before forwarding it.
In case of no-wait schedules, the forwarding operation must begin immediately once the entire frame has been received and processed by the bridge.
TSN generally supports frames buffering. 
However, keeping multiple frames in the same queue can lead to scheduling disruptions if the ordering is not guaranteed to be fixed or in case a frame is lost.
Therefore, frames are \emph{isolated} by storing at most one frame in a queue at any time.
No-wait schedules inherently adhere to frame isolation and use only a single queue per egress port.
Finally, a bridge cannot forward two frames via the same network link at once.
Thus, the schedule needs to guarantee \emph{mutual exclusive transmissions} for each frame.

All these constraints must be met in the gate control list of each bridge in order for a schedule to be valid.
Note that this does not make any claims about the quality of the schedule.
For example, rejecting all streams ($\mathcal{S} = \emptyset$) with an empty gate control list would be valid, but not helpful.
Therefore, we next discuss the optimization goal of the scheduling synthesis.

\subsection{Optimization Goal}
\label{sec:problem-statement:optimization-goal}

The primary optimization goal is to admit as many streams as possible, which is equivalent to minimizing the number of rejected streams.

\begin{equation}
	\min \left |\mathcal{S}^{rej}\right |
\end{equation}

In this work, we consider a dynamic system instead of performing a single offline computation. 
Hence, we try to minimize the number of rejected streams in each iteration, while upholding the quality of service guarantees for already admitted streams:

\begin{equation}
	\forall i \in \mathcal{I} \min \left |\mathcal{S}^{rej}_i\right |: \mathcal{S}_i\setminus\mathcal{S}^{del}_i \in \mathcal{S}_{i+1}
\end{equation}

$\mathcal{I}$ denotes the sequence of iterations.
Note that $\mathcal{I}$ is unknown to the scheduler and could theoretically be infinite.
Hence, as future iterations are unknown, the scheduler cannot purposefully reject streams in an iteration $i$ for the sake of admitting other streams in a later iteration $j$.

Finally, the runtime and resource requirements of the scheduler need to be low enough, so that even large-scale networks and bigger stream sets can be served. 
This can be achieved by reducing the variables and limiting the computation time for solver-based solutions, bounding the conflict graph for conflict-graph-based approaches \cite{Geppert2024}, or using computationally cheap heuristics.


\section{Multicast Partitioning}
\label{sec:multicast-partitioning}

A system that does not allow for multicast communication and instead uses multiple unicast communications would send many duplicate frames from the stream's origin.
This leads to unnecessary resource consumption, especially for large numbers of destinations, i.e., due to the increased bandwidth demand the network's link capacities are exhausted faster.
Therefore, it is desirable to send a frame only once and then duplicate it at the bridge if needed.

On the other extreme, every destination of a stream is served by the same tree-based multicast.
This leads to complex scheduling decisions to which the scheduler may not be able to find a solution anymore, as either queues or the downstream egress ports must be available at every bridge.
When using a no-wait scheduler the egress ports must be available.

The requirements are application specific with multiple parameters -- including the granularity of multicast partitioning on the spectrum between these two extremes -- that need to be considered.
Thus, we propose a method that enables partitioning on any granularity along this spectrum. 
This allows to save bandwidth through the use of multicasts, while maintaining easy schedulability of the streams and/or reducing the required queuing resources.
To be more precise, we suggest a preprocessing phase for multicast partitioning.
As shown in Figure~\ref{fig:two-step-process}, the multicast trees of the new streams in $\mathcal{S}^\text{add}$ are partitioned before handing them over to the scheduler for the scheduling synthesis. 
Partitioning of a multicast tree is performed when the distance in network hops between the destinations is above a threshold (cf. Section~\ref{sec:multicast-partitioning:partitioning-threshold}), i.e., the multicast tree is too widespread.
To uphold the quality of service for the application, either all partitions of a multicast tree must be scheduled or none at all.
Partial scheduling is not allowed.
In the following, we present the partitioning method in detail.

\subsection{Distance-based Grouping}

Our goal is to partition multicast trees into smaller multicast or even unicast trees, so that the schedule becomes easier to create, while the bandwidth benefits from using multicasts are preserved.
The destination group of a widespread multicast tree is split into multiple smaller trees based on the proximity of the destinations to each other in terms of network hops.
If a multicast tree has multiple destinations in close proximity to each other, these should remain in the same tree.
Since a frame sent to one of these destinations is already very close to the rest of the group, we expect the ensuing bandwidth savings to outweigh the increase in scheduling complexity.
In contrast, more remote destinations should form separate groups, in order to prevent the scheduling from being hindered by a branch of the multicast tree at the other side of the network.
We call the maximum hop distance a destination may have to the closest member in its group the \emph{partitioning threshold} (cf. Section\ref{sec:multicast-partitioning:partitioning-threshold}).
It is a parameter that must be provided by the user.

Note that other distance measures would also be possible.
For example, a metric based on link transmission times might be more suitable for heterogeneous networks.
However, using a hop-based distance metric has the advantage of being easily extractable from the topology information of the network, while providing a good estimation of the distances within the network.

\begin{algorithm}[t]
	\KwIn{Multicast stream $m$, threshold $t$}
	\KwOut{Destination groups $\mathcal{DG}$}
	\tcp{Compute distances between all destinations and source of m}
	$D \leftarrow apsp(m)$\; 
	$G \leftarrow \{\argmax_{\forall x \in m_{dst}} D(m_{src}, x)\}$\;
	\While{$\exists~dst \in m_{dst} \notin \mathcal{DG}$}{
		$dst \leftarrow \argmin_{\forall y \notin DG, \forall z \in G} D(y, z)$\;
		\If(add closest destination){$\min_{\forall z \in G} D(dst,z) \leq t$}{
			$G \leftarrow G \cup {dst}$\;
		}
		\Else(create new destination group){
			$\mathcal{DG}$.append($G$)\;
			$G \leftarrow \{dst\}$\;
		}
	}
	\Return{$\mathcal{DG}$}\;
	\caption{Multicast Partitioning Algorithm}
	\label{alg:splitting}
\end{algorithm} 

Algorithm~\ref{alg:splitting} provides the pseudocode for the partitioning procedure of a multicast stream $m$.
First, the shortest paths between all destinations as well as the source need to be computed (all-pair-shortest-path, APSP).
This allows us to access the distance ($D$) between any two nodes.
Next, the destination with the largest distance to the source of $m$ is selected to form the first group $G$.
Iteratively, the next closest destination $dst$ to any element in G is determined.
In case the distance of $dst$ from G does not exceed the given partitioning threshold $t$, it is appended to $G$.
Otherwise, the group is complete since no further destination exists with a distance to G below or equal to the threshold.
Thus, a new group is initialized with $dst$ and the procedure is repeated with the new group.
Once all destinations are processed, all groups are returned.

\subsection{Partitioning Threshold Parameter}
\label{sec:multicast-partitioning:partitioning-threshold}

The success of the multicast partitioning strongly depends on the partitioning threshold parameter.

The partitioning threshold covers the spectrum between a single multicast tree and a set of unicast communications.
Setting it to zero results in full partitioning with unicast streams only.
Setting it larger or equal to the network's diameter leads to no partitioning.
The threshold needs to be selected depending on the network environment.
For instance, in a network with a diameter of 6 hops, destinations within a proximity of 4 hops should not be considered close, while in a network with diameter 10, they could be considered close.

We saw good results using thresholds around 4 for networks with 49 bridges.
Note that 2 hops out of these 4 hops are already needed to connect the end devices to the network, i.e., two end devices connected to the same bridge have a distance of 2.
As the network size and thus the distances between source and destinations grows, the threshold needs to be increased.
Additionally, the threshold needs to be adapted to the scheduling algorithm (cf. Section~\ref{sec:eval:threshold}).
In order to account for the additional complexity of dynamic environments, an interesting avenue for future research might therefore be the investigation of a mechanism for the dynamic adaptation of the threshold parameter and possibly the partitioning of already scheduled multicast streams.


\section{Evaluation}
\label{sec:eval}

In the following, we first describe our evaluation setup before discussing the scheduling algorithms that we used in our evaluation and finally, present our results.

\subsection{Setup}

We analyzed the schedules using a Linux machine with two AMD EPYC 7401 24-core processors.
The execution was limited to 16 cores for the CP, 8 cores for the GFH scheduling, and 1 core for Greedy, based on the capability and benefit of parallelization in preliminary evaluations.
The machine was equipped with \SI{256}{\giga\byte} RAM, although much less RAM would have sufficed.

We considered \SI{1}{\micro\second} for the propagation delay and \SI{4}{\micro\second} for the processing delay in a network with \SI{1}{\giga\bit\per\second} links.
The frame sizes were randomly selected from \{125, 250, 500, 750, 1000, 1500\}\unit{\byte} and the periods from \{250, 500, 1000, 2000\}\unit{\micro\second}.
We used a macrotick of \SI{1}{\micro\second}, i.e., schedules can be defined with a granularity of \SI{1}{\micro\second}.
The number of destinations for a stream was selected from \{1, 2, 4, 8, 16\} for networks with 49 bridges or from \{1, 2, 4\} for networks with only 12 bridges such that $\sim$50$\%$ of the streams communicated with a single destination (unicast), $\sim$25$\%$ with 2 destinations (multicast), etc.
The largest group (16 or 4 destinations) was equal to the second-largest group.
This results in an average number of destinations per stream of $\sim$3 or $\sim$2, which aligns with other multicast scheduling evaluations \cite{Schweissguth2020,Li2021}, while also accounting for larger destination sets.

We evaluate the configurations based on three metrics: the number of rejected streams, the runtime, and the accumulated network throughput.

For the number of rejected streams, rejecting a \enquote{hard to schedule} stream is considered equivalent to rejecting an \enquote{easy to schedule} stream \cite{Geppert2025}.
Factors that make a stream harder to schedule include higher number of destinations, shorter period, and a larger frame size.
To this end, we additionally measure the accumulated network throughput, which is less prone to be biased towards \enquote{easy to schedule} streams and thus helps us create a more comprehensive picture of the effectiveness of the synthesized schedules.
The accumulated network throughput represents the total amount of bits that arrive at their stream destination:

\begin{equation}
	\max \sum_{s \in \mathcal{S}} \frac{s_\text{frame size} \times \left |s_\text{destinations}\right |}{s_\text{period}}
\end{equation}

\subsection{Scheduling Algorithms}

To determine the effectiveness of our multicast partitioning, both phases of the workflow must be completed, i.e., the partitioned streams must be forwarded to a scheduling algorithm.
Since the partitioning itself is agnostic of a particular scheduling algorithm, we run the evaluations with three different schedulers: a constraint-programming-based  no-wait approach (CP-NW), the Greedy strategy from Raagaard et al. \cite{Raagaard2017}, and the conflict-graph-based GFH scheduling approach from Falk et al. \cite{Falk2022}.

The constraint programming approach\footnote{\url{https://github.com/gepperho/CP-multicast-partitioning-and-scheduling}} was implemented in python using IBM's CPLEX solver.
The no-wait principle reduces the considered solution space, which decreases the schedulability \cite{Vlk2022}.
However, the solver might find a solution faster in a reduced solution space \cite{Xue2024}.
In case a bridge needs to duplicate a frame to forward it via two different egress ports, both duplicates adhere to the no-wait principle.

The Greedy scheduling is a very fast strategy that iteratively inserts streams into the schedule based on an as-soon-as-possible principle.
It allows for queuing and uses multiple egress queues so that frames can overtake each other.
Hence, duplicate frames can be forwarded at different times.
The original version uses application graphs to model precedence constraints between streams.
For the sake of simplicity and as we only use the scheduler to test our partitioning approach, we omitted this additional feature in our implementation.

Finally, we implemented a slightly adapted version of the GFH scheduling, so that it also allows for multicast scheduling\footnote{\url{https://github.com/gepperho/ConflictGraphBasedScheduling/tree/multicast_implementation}}.
We updated the configurations (vertices) in the conflict graph to represent multicast trees instead of unicast routes.
Analogous to the CP scheduler, frame duplicates are all forwarded according the no-wait principle.
The conflict graph is solved by the Greedy Flow Heap heuristic \cite{Falk2022}.
We implemented Greedy and GFH in C++ and compiled them with GCC~13.

To create the multicast trees, we used an intermediate node distance algorithm \cite{Nass2023}.
The algorithm takes the shortest path from the source to one destination, and then iteratively connects the remaining destinations via the shortest path to the existing tree.
If two equally short paths exist, the destination is attached to the node closer to the root.
This results in a low total number of links in the multicast tree, although at the cost of a potentially higher tree depth.

\subsection{Results}

In the following, we first present the results of multicast partitioning on different network topologies.
Afterwards, we discuss how the different scheduling algorithms are affected by the partitioning, followed by an analysis of the threshold parameter.
Finally, we summarize the findings.

\subsubsection{Network Topology}

\begin{figure*}
	\captionsetup[subfigure]{width=0.9\textwidth}
	\centering
	\begin{subfigure}[t]{0.49\textwidth}
		\centering
		\includegraphics[width=0.8\linewidth,alt={Line plot showing the difference in rejected streams when using the multicast partitioning on different network topologies. In most displayed cases the multicast partitioning lead to fewer rejected streams. The presented network topologies are: grid, random, waxman, preferential attachment and small world.}]{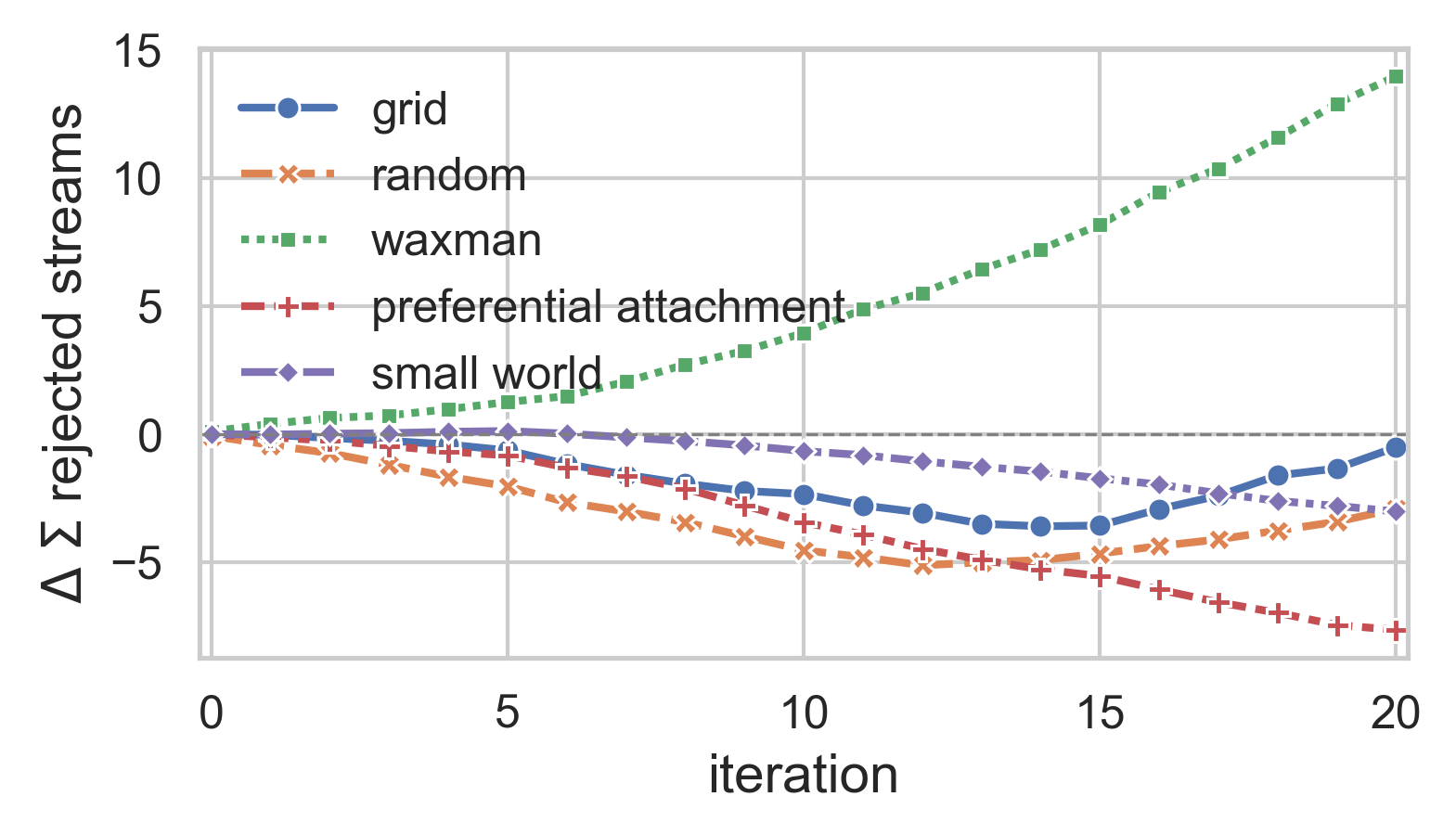}
		\caption{Difference in cumulative rejected streams. Negative values imply that the partitioning approach rejected fewer streams.}
		\label{fig:delta:rejected-clusters}
	\end{subfigure}
	\begin{subfigure}[t]{0.49\textwidth}
		\centering
		\includegraphics[width=0.8\linewidth, alt={Line plot showing the difference in accumulated network throughput when using the multicast partitioning on different network topologies. In most displayed cases the multicast partitioning leads to a higher network throughput. The presented network topologies are: grid, random, waxman, preferential attachment and small world.}]{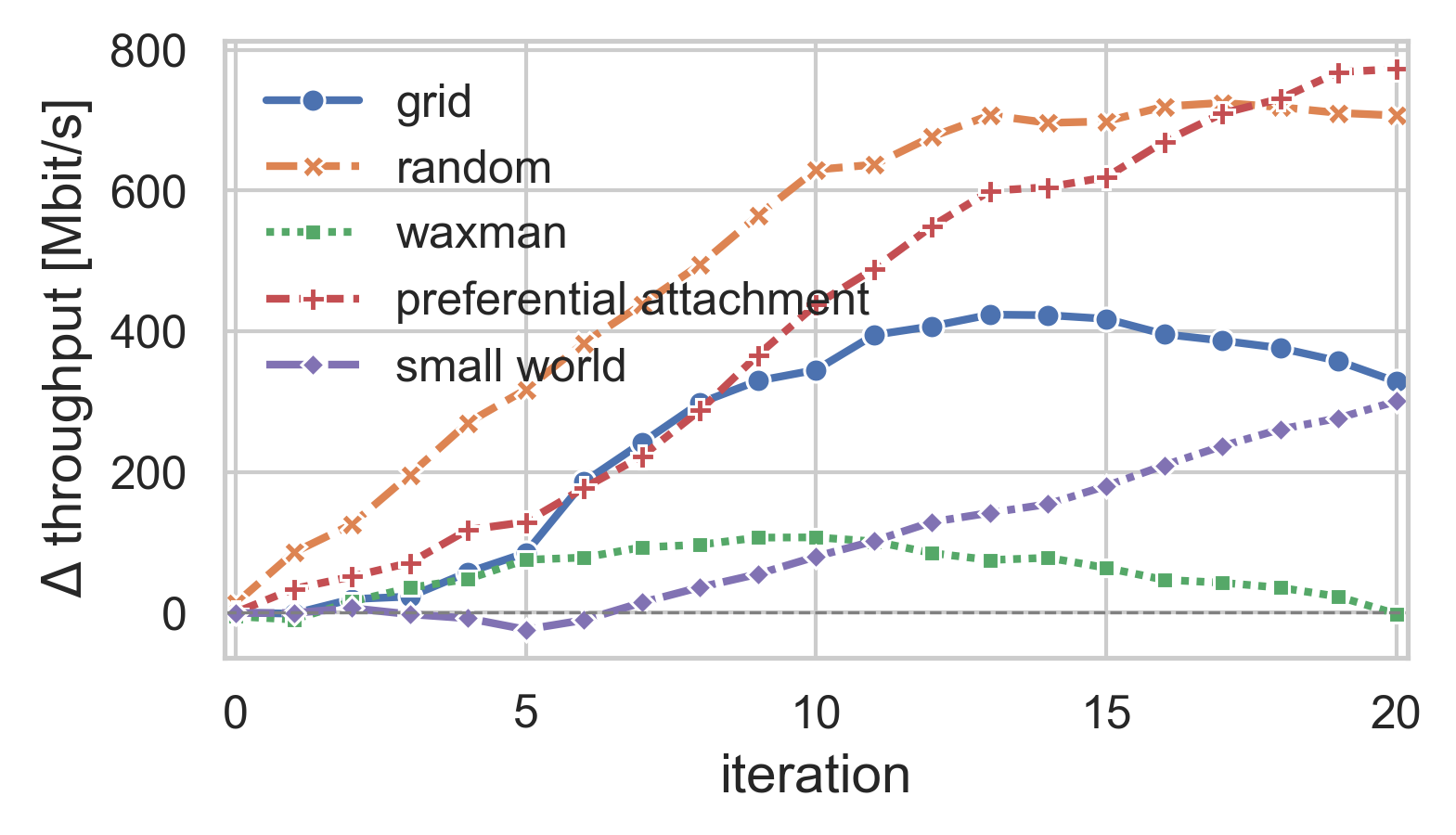}
		\caption{Difference in admitted network throughput. Positive values imply that the partitioning approach had a higher admitted throughput.}
		\label{fig:delta:accepted-traffic}
	\end{subfigure}
	\caption{Comparison of multicast partitioning versus non-partitioning on different network topologies.}
	\label{fig:delta}
\end{figure*}

The network topology is one of the most important factors impacting the success of our partitioning approach.
To present a broad range of topology properties, we investigated the following topologies: grid, Erd\H{o}s--R\'enyi (random), Waxman, Barab\'asi--Albert (preferential attachment), and small world \cite{Newman2018}.
For each topology, we used the GFH scheduler to schedule between 100 and 250 streams in the first iteration.
In each later iteration 40 streams were added to the system and 20 streams left it.
An iteration denotes one, possibly major, update to the traffic plan, which represents the system changes over time.
Thereby, an iteration can denote an arbitrary, flexible time interval, i.e., defined by the application needs.
Each topology was formed by 49 bridges and each bridge was connected to a single end device.
We evaluated the partitioning approach with thresholds of 3-5 for each topology.
In the following, we always present the results that were obtained with the threshold resulting in better outcome for the respective topology.
Note that the number of streams vary between the topologies as some topologies can admit more streams than others, and we wanted them all to be highly utilized while not being completely overcrowded.

Figure~\ref{fig:delta} displays the difference (delta) between multicast partitioning and non-partitioning in terms of the sum of rejected streams and throughput.
The Waxman topology was the only one where the partitioning rejected constantly more streams. 
However, the network throughput was higher than in the non-partitioning case.
The partitioning performed exceptionally well on scale-free networks, namely Erd\H{o}s--R\'enyi (random) and Barab\'asi--Albert (preferential attachment) graphs.
Since there was a smaller benefit in rejected streams and throughput on small world graphs, the good performance on the scale-free graphs does not seem to result from the small world effect (node pairs are connected by short paths \cite{Newman2018}).
Instead, we attribute it to the power-law degree distribution, which leads to well-connected hub nodes and highly utilized links between them.
For the random and grid networks, the number of additionally rejected streams without partitioning grows first and then starts to shrink again.
This trend occurs because we look at highly congested networks, where the partitioning approaches cannot schedule more streams at some point, because the network load has practical limits, while the non-partitioning approaches have capacity left from earlier rejections and catch up.
With more iterations, it is very likely that the same behavior would be seen for other topologies as well.
Yet, we see that the partitioning approach utilizes the available resources better.
In terms of runtime, all schedules were created in less than one second (up to \SI{0.8}{\second} for the initial iteration, and about \SI{0.6}{\second} for iteration 20).

\subsubsection{Impact on Scheduling}

We now take a closer look at the impact of the multicast partitioning on different scheduling approaches.
We first investigated the two heuristic schedulers, i.e., GFH and Greedy, on Erd\H{o}s--R\'enyi random networks with 49 bridges.
In a second step, we evaluated CP-based scheduling on a smaller network with 12 bridges.


\begin{figure*}
	\centering
	\captionsetup[subfigure]{width=0.9\textwidth}
	\begin{subfigure}[t]{0.32\textwidth}
		\centering
		\includegraphics[width=1\linewidth, alt={Line plot showing the GFH heuristic with and without multicast partitioning in terms of rejected streams over the course of several iterations. The configuration with multicast partitioning rejects fewer streams (threshold=4).}]{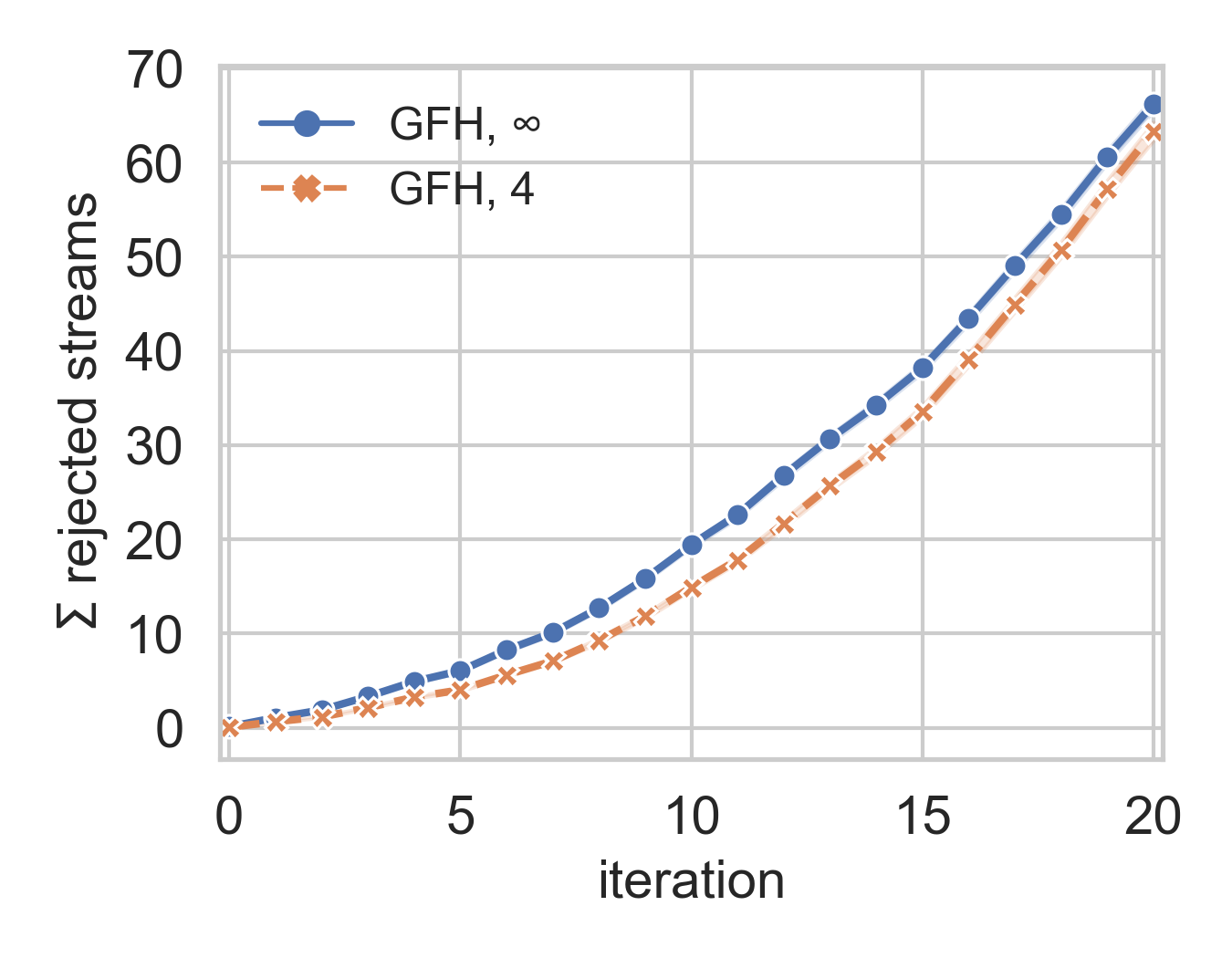}
		\caption{Cumulative rejected streams using GFH scheduling.}
		\label{fig:49:rejected-gfh}
	\end{subfigure}
	\begin{subfigure}[t]{0.32\textwidth}
		\centering
		\includegraphics[width=1\linewidth, alt={Line plot showing the Greedy scheduling with and without multicast partitioning in terms of rejected streams over the course of several iterations. The configuration with multicast partitioning rejects fewer streams (threshold=3). The Greedy scheduling rejects much more streams than GFH.}]{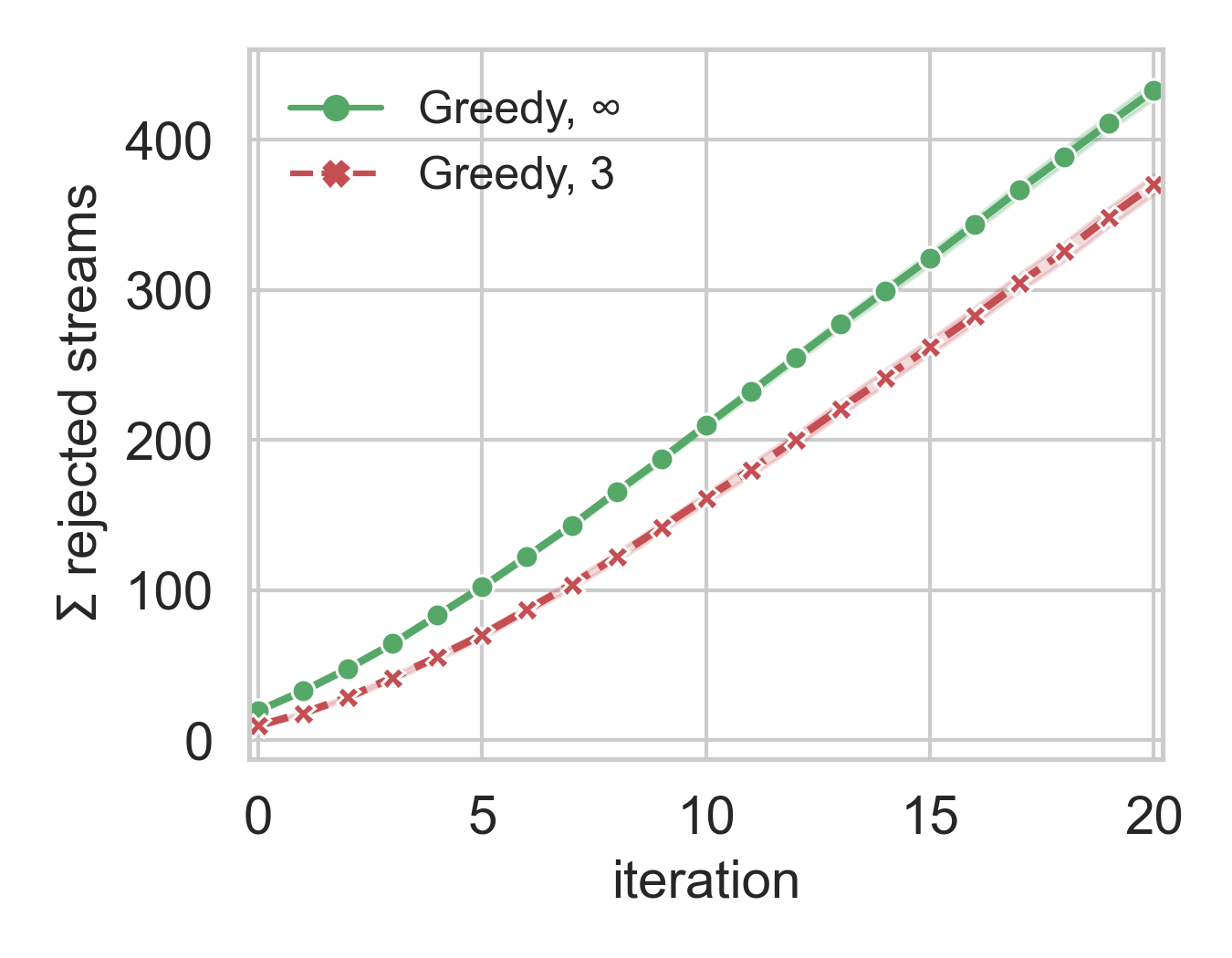}
		\caption{Cumulative rejected streams using Greedy.}
		\label{fig:49:rejected-greedy}
	\end{subfigure}
	\begin{subfigure}[t]{0.32\textwidth}
		\centering
		\includegraphics[width=1\linewidth, alt={Line plot showing the GFH and Greedy schedulings performance in terms of accumulated network throughput with and without multicast partitioning over the course of several iterations. Multicast partitioning leads to higher throughput in the shown examples. Further, the throughput of GFH is much higher than the one of the Greedy Scheduling.}]{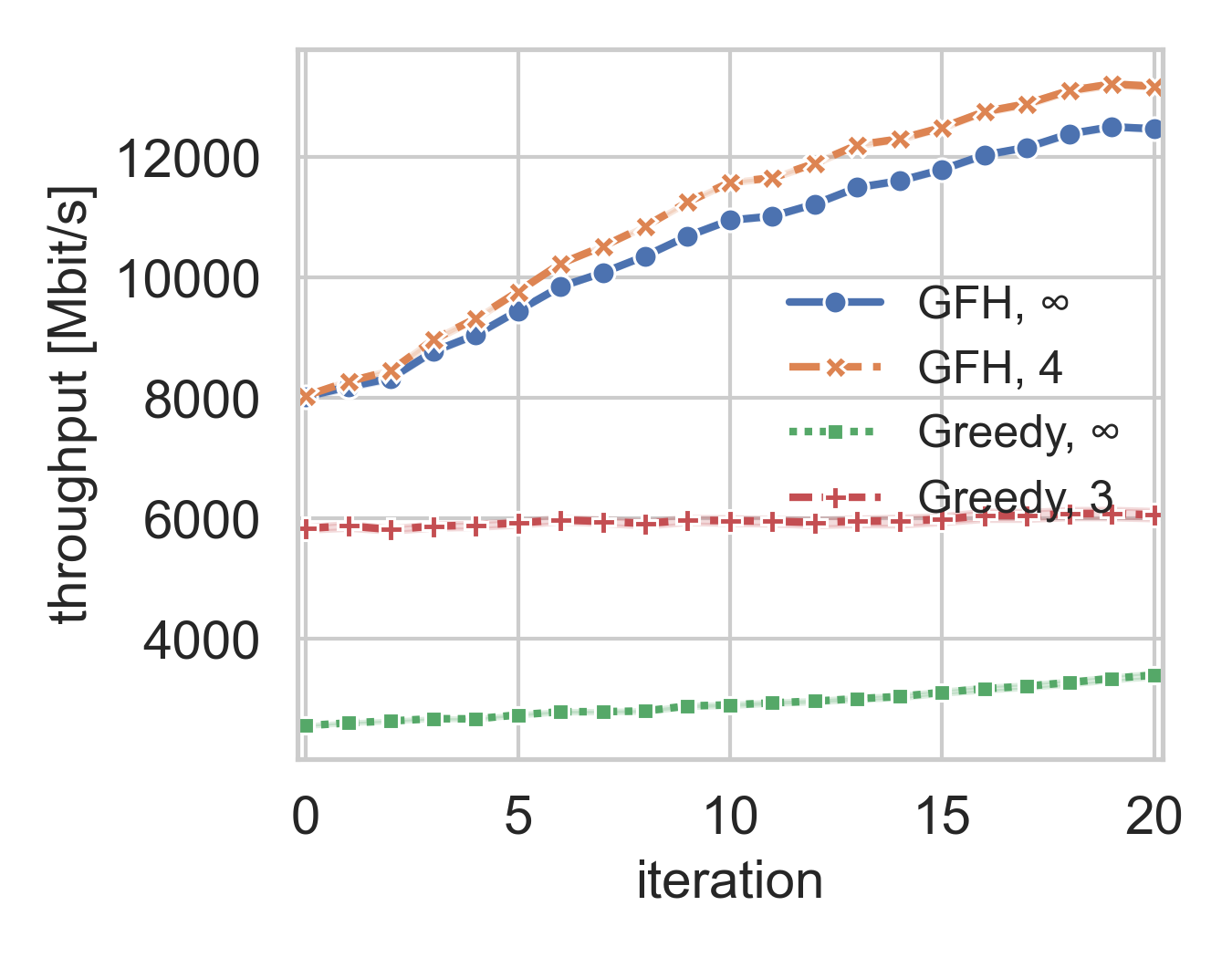}
		\caption{Accumulated admitted network throughput for Greedy and GFH.}
		\label{fig:49:traffic}
	\end{subfigure}
	\caption{Evaluation of GFH and Greedy heuristics on random networks with 49 bridges after either performing or not performing multicast partitioning. The number after the scheduler name denotes the partitioning threshold.}
	\label{fig:49}
	
\end{figure*}

For the evaluation of the heuristics, 250 streams were added in the initial iteration.
40 streams were added and 20 removed in each subsequent iteration.
Partitioning was best for GFH with a threshold of 4, and with a threshold of 3 for Greedy.
We display the cumulative number of rejected streams in Figures \ref{fig:49:rejected-gfh} and \ref{fig:49:rejected-greedy}, and the network throughput in Figure~\ref{fig:49:traffic}.	
As we can see, in both cases, the number of rejected streams is smaller when using multicast partitioning compared to non-partitioning.
In the last iteration, the difference is about 5\% for GFH and about 15\% for Greedy.
In both cases, the relative difference is higher in earlier iterations (up to 50\% difference).
Additionally, the admitted network throughput was higher with both schedulers when applying the partitioning (up to 5\% with GFH, up to 125\% with Greedy). 
Note that poorly chosen thresholds can lead to more rejected streams compared to the non-partitioning.
However, even in such cases, we observed a higher admitted throughput with the multicast partitioning.
With regard to runtime, GFH stayed below \SI{1}{\second} and Greedy below \SI{0.1}{\second}.
Further, the partitioning led to a small runtime reduction of the GFH scheduler and a small increase for Greedy.
This shows that the simplified conflict graph creation and independent set computation resulting from the multicast partitioning indeed benefits the efficiency of GFH.
The Greedy scheduler, in contrast, becomes less efficient under the increased load from the higher number of multicast/unicast streams.
The results were less pronounced when using non-harmonic periods.

\begin{figure}
	\centering
	\begin{minipage}{.48\textwidth}
		\centering
		\includegraphics[width=0.7\linewidth]{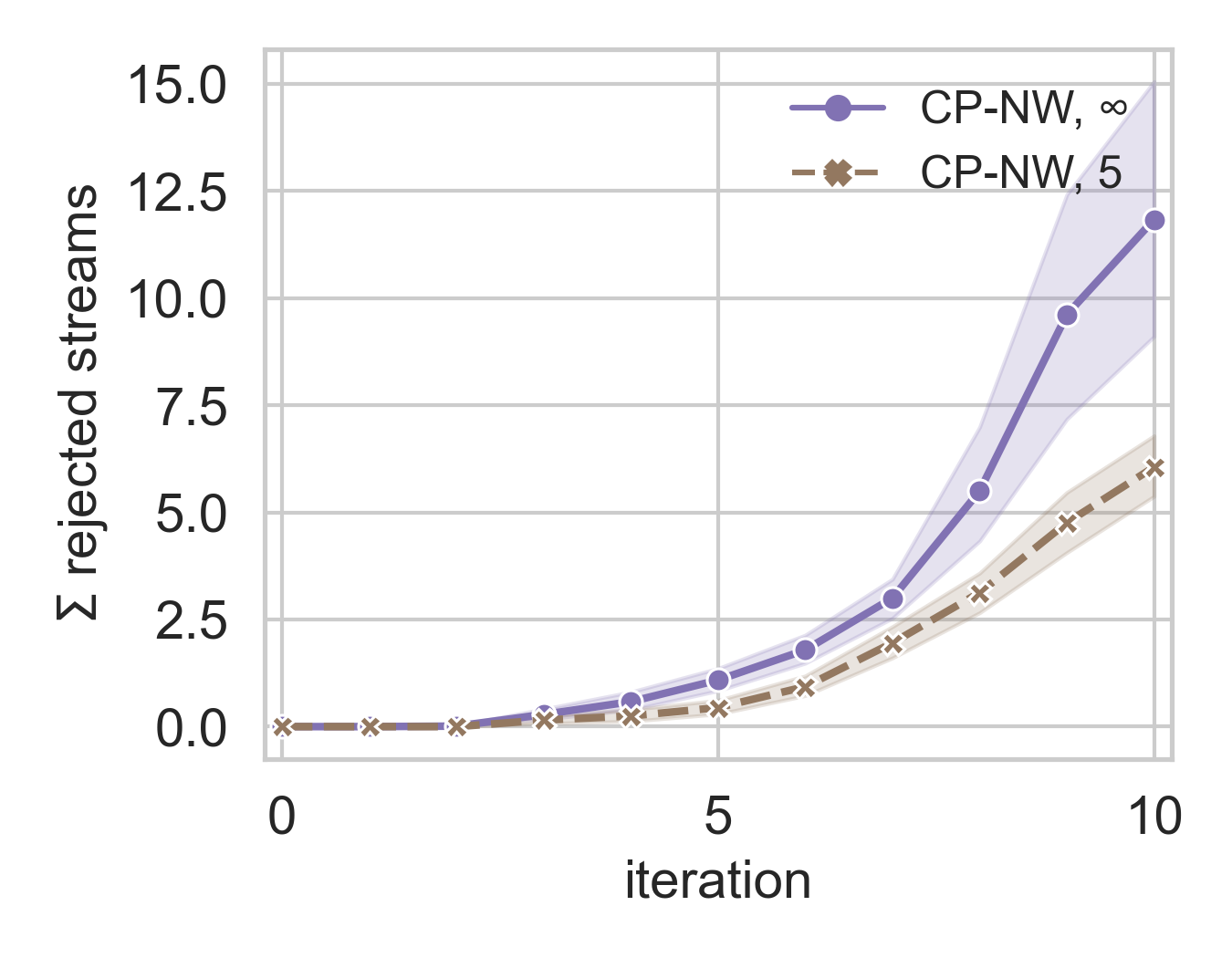}
		\caption{Cumulative rejected streams using a no-wait constraint programming scheduler with a scheduling time limit of \SI{60}{\second}.}
		\label{fig:cp:rejected-streams}
	\end{minipage}%
	\hfill
	\begin{minipage}{.48\textwidth}
		\centering
		\includegraphics[width=0.7\linewidth]{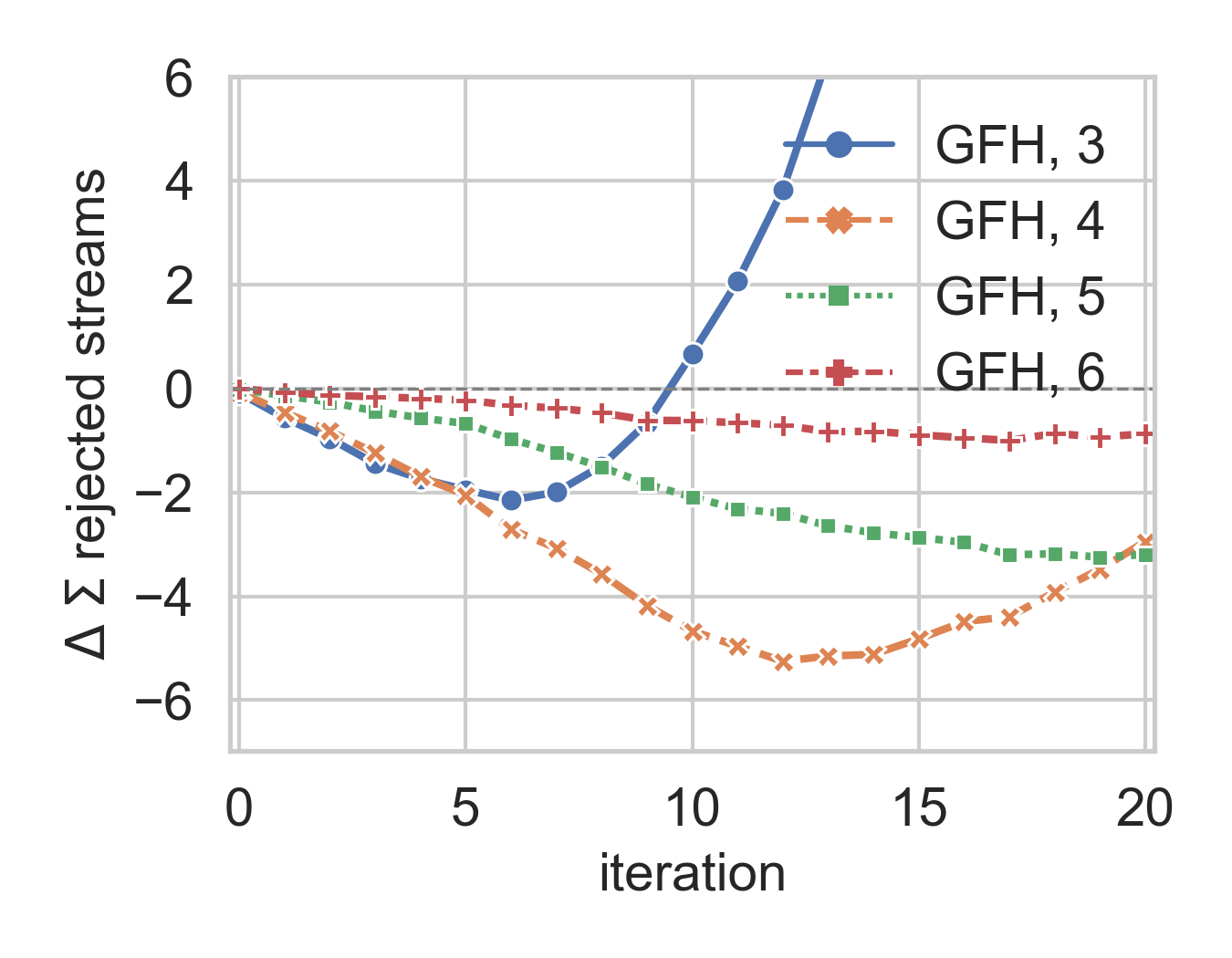}
		\caption{Difference in cumulative rejected streams.  Negative values imply that the GFH scheduling with the given partitioning threshold rejected fewer streams than GFH scheduling without partitioning.}
		\label{fig:param:deltagfhcumulative-rejected-clusters}
	\end{minipage}
\end{figure}

Next, we examined the multicast partitioning on a CP-based scheduler.
Since solver-based schedulers have much higher runtimes and do not scale as well as heuristic solutions, we reduced the network size to 12 bridges.
In the initial iteration, 150 streams were scheduled.
In each subsequent iteration, 20 new streams were added, and 10 streams left the system.
Note that we only used 10 iterations here, as the CP approach runs longer than the heuristic approaches, and the smaller scenario length accommodates the smaller network well, i.e., the differences are well visible after 10 iterations already.
We allowed for a maximum scheduling time of 60 seconds per iteration, which sufficed in most, but not all, experiments.
Figure~\ref{fig:cp:rejected-streams} visualizes the sum of cumulative rejected streams with and without the multicast partitioning.
In later iterations, partitioning (with a threshold of 5) performs better, i.e., fewer streams are rejected than with the non-partitioning.
Network throughput was higher when partitioning was applied due to the lower rejection rates, although the difference was small.
When increasing the time limit, the difference in the number of rejected streams vanishes, although the multicast partitioning approach still had a slightly higher throughput in the later iterations.
This behavior matches the results that we observed for the heuristics.


\subsubsection{Partitioning Threshold}
\label{sec:eval:threshold}


Finally, we investigated the impact of the partitioning threshold, using GFH as a representative scheduler.
Figure~\ref{fig:param:deltagfhcumulative-rejected-clusters} shows the difference in cumulative rejected streams when applying partitioning with thresholds ranging from 3 to 6 relative to non-partitioning.

Remember that, in our topologies, every bridge is connected to only one end device.
Since two hops are already required for the link from and to the end device, a threshold of 3 thus means that two destinations are maintained in the same multicast tree only if their adjacent bridges are neighbors.
We therefore only evaluate thresholds equal to or above 3.

As we can see in Figure~\ref{fig:param:deltagfhcumulative-rejected-clusters}, partitioning with any of our tested thresholds led to a rejection of fewer streams initially, although the difference is negligible when using a threshold of 5 or 6.
With a threshold of 3, the schedulability deteriorates after 6 iterations, and after 10 iterations, more streams are rejected than non-partitioning (positive delta values).
We assume that this effect is caused by the increased bandwidth requirements, as more and more additional unicast or multicast streams overcrowd the network.
We observed the best results with a threshold of 4.
However, it should be pointed out that optimal value for the threshold is dependent on many influences such as the network size or the employed scheduling algorithm.
For example, Greedy achieved the best results with a threshold of 3.
As this scheduler rejects much more streams overall, the network does not as easily become saturated as with the GFH scheduler.
Therefore, bandwidth does not as quickly turn into a bottleneck.

\subsubsection{Summary}
We have seen that multicast partitioning can improve the scheduling results in terms of rejected streams and accumulated network throughput substantially. 
These improvements come from a reduction of scheduling conflicts, because each smaller multicast tree has fewer overlaps with other trees, leading to fewer dependencies when creating the schedule.
In addition, the multicast partitioning is not bound to a specific scheduling approach.
In our experience, the only requirement placed on the scheduler is the implementation of a no-wait principle or similar method like as-soon-as-possible placement.
The impact of the multicast partitioning depends on two factors: 
First, on the network topology, whereby scale-free topologies seem to benefit most from partitioning.
Second, on the partitioning threshold, which is specific to the applied scheduling algorithm with values ranging between 3 and 4 in our experiments.


\section{Conclusion}
\label{sec:conclusion}

Large multicast trees can make the scheduling in time-sensitive networks harder.
In this paper, we therefore investigated a multicast partitioning in conjunction with conventional scheduling approaches.
The multicast partitioning divides widespread multicast trees into smaller multicast or unicast trees, based on the distance between the leaf nodes.
The degree of the partitioning, i.e., where to partition, can be configured by a user-defined partitioning threshold.
With the partitioning, we obtained improved scheduling results in terms of fewer rejected streams, higher admitted accumulated network throughput, and for some schedulers shorter runtimes.
The results were consistent for different network topologies and scheduling approaches.

Future work should focus on dynamic partitioning threshold, so that the threshold is automatically adapted to the network size, topology, and utilization.
This should also include an adaptation of the threshold while the system is active, in case the system would benefit from a different threshold.

\section*{Acknowledgments}
Special Thanks to Melanie Heck for the valuable input on the language and presentation of the paper.

	\bibliographystyle{abbrvnat}
	\bibliography{graph_related_literature.bib}
	
\end{document}